\documentstyle[mprocl]{article}

\input{psfig.sty}
\def\Journal#1#2#3#4{{#1} {\bf #2}, #3 (#4)}

\def\PRL{\em Phys. Rev. Lett.}
\def\PRD{{\em Phys. Rev.} D}

\def\be{\begin{equation}}
\def\ee{\end{equation}}
\def\bea{\begin{eqnarray}}
\def\eea{\end{eqnarray}}

\begin{document}

\title{Relativistic disks as sources of Kerr-Newman fields}
\date{}

\author{Tom\' a\v s Ledvinka, Martin \v Zofka and Ji\v r\'\i \  Bi\v c\'ak}

\address{
{\footnotesize Department of Theoretical Physics, Faculty of Mathematics and Physics,}\\
{\footnotesize Charles University, V Hole\v sovi\v ck\'ach 2, 180 00 Prague 8, Czech Republic}\\
{\scriptsize ledvinka@mbox.troja.mff.cuni.cz,
zofka@otokar.troja.mff.cuni.cz,
bicak@mbox.troja.mff.cuni.cz}
}


\maketitle
\abstracts{
If an appropriate region of Kerr-Newman space-time
is removed and suitable identifications are made,
the resulting space-time can be interpreted as an infinitely thin disk
producing the original  electromagnetic and gravitational fields.
We choose the shape of the regions removed in such a way that radial
pressures in the disks vanish.
Even the very inner parts  such as ergoregions may exist around the disks.
The surface energy-momentum tensor of the disks
is checked to satisfy the weak and strong energy conditions. To emphasize
the reality of these sources
two models of the disks are presented: (i) rotating conductive charged
rings which are supported against collapse by their internal pressure,
(ii) two counter-rotating streams of charged  particles
moving along circular electro-geodesics.
All these disk sources form a three-dimensional parameter space
with specific electric charge $Q/M$, angular momentum $a/M$
and the size of the excluded region being the parameters.
}

\def\boxit#1{\vbox {\hrule \hbox{\strut \vrule{} #1 \vrule}\hrule}}
\section{Introduction}
Although much effort has been devoted to discovering exact
solutions of Einstein's equations, there are only a few
"physically acceptable" solutions available.
What is in particular lacking are sources which would produce
numerous known vacuum or electrovacuum metrics burdened by
singularities.

Most vacuum {\bf static} Weyl solutions can arise as the metrics of
counter-rotating relativistic disks\cite{blbk,blbp}.
The method used to construct such sources resembles the method of
images in electrostatics. Alternatively, it consists
in first cutting out a portion of a stationary space-time
"between" two hypersurfaces
and then glueing suitably these hypersurfaces together. The jump in
the normal derivatives of the resulting potentials induces a
matter distribution in the disk which arises due to the
identification. One has then to analyze whether the matter is
physically acceptable.

Later the method has been used to
construct physical disk sources of vacuum {\bf stationary}
space-times, in particular, of Kerr metrics\cite{bl,lbp}.

A similar procedure can be used to construct disk sources of
stationary electrovacuum space-times as, for example, of Kerr-Newman fields.
The resulting disks then contain rotating matter and also electric
currents which arise due to the jumps of the (normal components of)
electromagnetic fields.

\section{Kerr-Newman Fields}
It is convenient  to use the
Weyl-Papapetrou (W-P) form of the line element
\be
ds^2 = e^{-2\nu}\left[ e^{2\zeta}(d\varrho^2+dz^2) +
       \varrho^2 d\varphi^2 \right]-e^{2\nu}(dt+A d\varphi)^2 ,
\ee
with the simple relation to Boyer-Lindquist (B-L) coordinates
\be
z = (r-M)\cos \vartheta, \quad
\varrho = \sqrt{r^2-2Mr+a^2+Q^2} \sin \vartheta.
\ee
Regarding these relations, functions $\nu,\zeta,A$ entering
the metric can be found from the standard form of the K-N metric
in B-L coordinates.

In the W-P coordinates we exclude the region $z\in[-b,b]$.
Using then the formalism for thin shells\cite{baiz} which can
carry an electric charge\cite{ku}, we find that
the disk has no radial pressure and the surface stress-energy
tensor and electric current can simply be expressed
in terms of the derivatives of the metric and electromagnetic
potentials at $z=b$ (indices $a$,$b$ correspond to $t$,$\varphi$)
\be{ S}_{ab} = {\sqrt{g_{\varrho\varrho}}\over 8 \pi} \left( {{ g}_{ab}\over g_{\varrho\varrho}}\right)_{,z} , \ee
\be{ J}_{a} = -{1 \over 2\pi}\quad \left( { A}_a \right)_{,z} . \ee
The existence of dragging results in a non-diagonal stress-energy
tensor. The following two models of the
disk are considered.
In the first model the disk is constructed from rings with
internal pressure. The necessary condition is that
${ S}_{ab}$ can be diagonalized, i.e.,
one can find, at any radius, $\varphi$-isotropic observer
(FIO) who sees only energy density and azimuthal
pressure.  In the charged case the rings are also
charged and conduct a current to form the current density ${ J}_a$.

In the second model the disks are described as
two counter-rotating streams of freely moving charged particles.
The fact that in the static case their velocities
must be identical and opposite, allows one to find
$V_{\pm} = \pm \sqrt{S^\varphi_\varphi/S^t_t}$.
When dragging is present,
the stream velocities $U^a_\pm$ must be determined from
the geodesic (or the electro-geodesic) equation
\be {1\over 2} \epsilon_{\pm}  g_{ab,(\varrho)} U^a_\pm U^b_\pm =
            - \sigma_{\pm} F_{(\varrho) c} U^c_\pm . \ee
Although ${ S}^{ab}$ has three different
components, the Einstein and contracted Gauss-Codazzi equations
guarantee that the following decompositions
can be made
\be S^{ab} = \epsilon_+ U^a_+ U^b_+  +
                   \epsilon_- U^a_- U^b_- , \ee
\be J^{\alpha} = \sigma_+ U^a_+ +
             \sigma_- U^a_-. \ee
\section{Properties of the disks}
The central mass densities and the central electric charge
densities of the disks can be expressed in analytic forms
\be \epsilon_c = {{{M}\over{2\pi}}}{{(b+M)^2-a^2-Q^2(1+b/M)}\over{
 [(b+M)^2+a^2]^{3/2}[b^2+a^2-M^2+Q^2]^{1/2}}} , \ee
\be \sigma_c = {Q \over {2 \pi}}
 {{(b+M)^2-a^2}\over\left[{(b+M)^2+a^2}\right]^2}
 \sqrt{(b+M)^2+a^2\over b^2+a^2+Q^2-M^2} .
\ee
Other physical properties of the disks can be best exhibited
graphically. Here we limit ourselves to present just one example
corresponding to the disk producing the Kerr-Newman gravitational
and electromagnetic fields with $M=1$, $a=0.4$ and $Q=0.1$. In
the figures the disks with $b/M=0.912, 1.0, 1.1, 1.2, 1.3, 1.4,
1.5$ are considered.
The disk with $b/M=0.912$ cannot be
constructed from the counter-rotating streams,
even though it satisfies both the weak and strong energy conditions.
The disks become highly relativistic in central
regions but they all have "classical" properties at large $R$
(proper circumferential radius). The mass
and charge densities (as measured by FIOs)
decrease rapidly with $R$.
The disks have infinite extensions, their
total mass, charge, and angular momentum are finite. They should
have properties common with finite relativistic disks. And
they represent the only physical sources of Kerr-Newman
fields available today.

\begin{tabbing}
a \hskip6cm a\=a \hskip6cm a\kill
\psfig{file=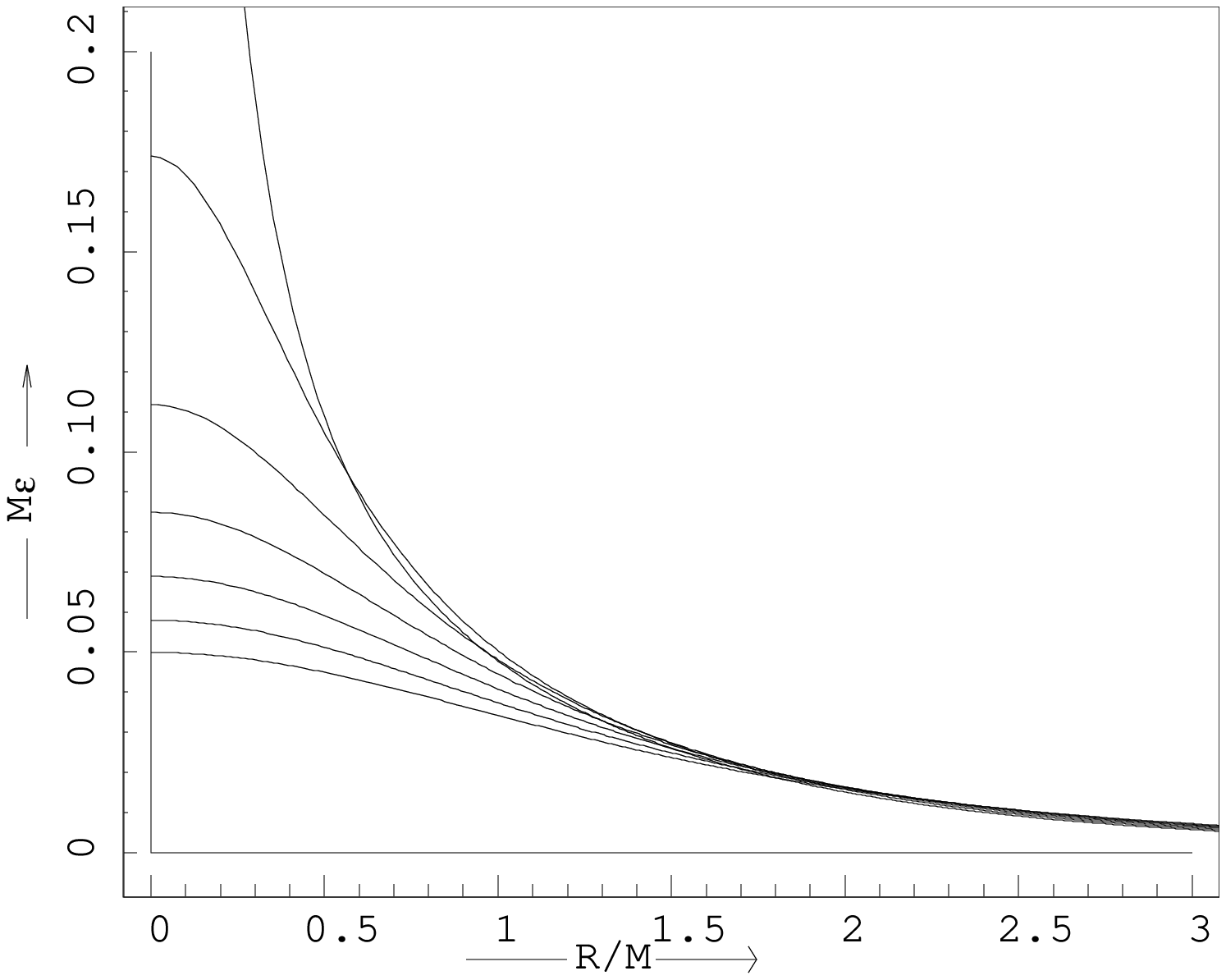,width=5.5cm}\>
\psfig{file=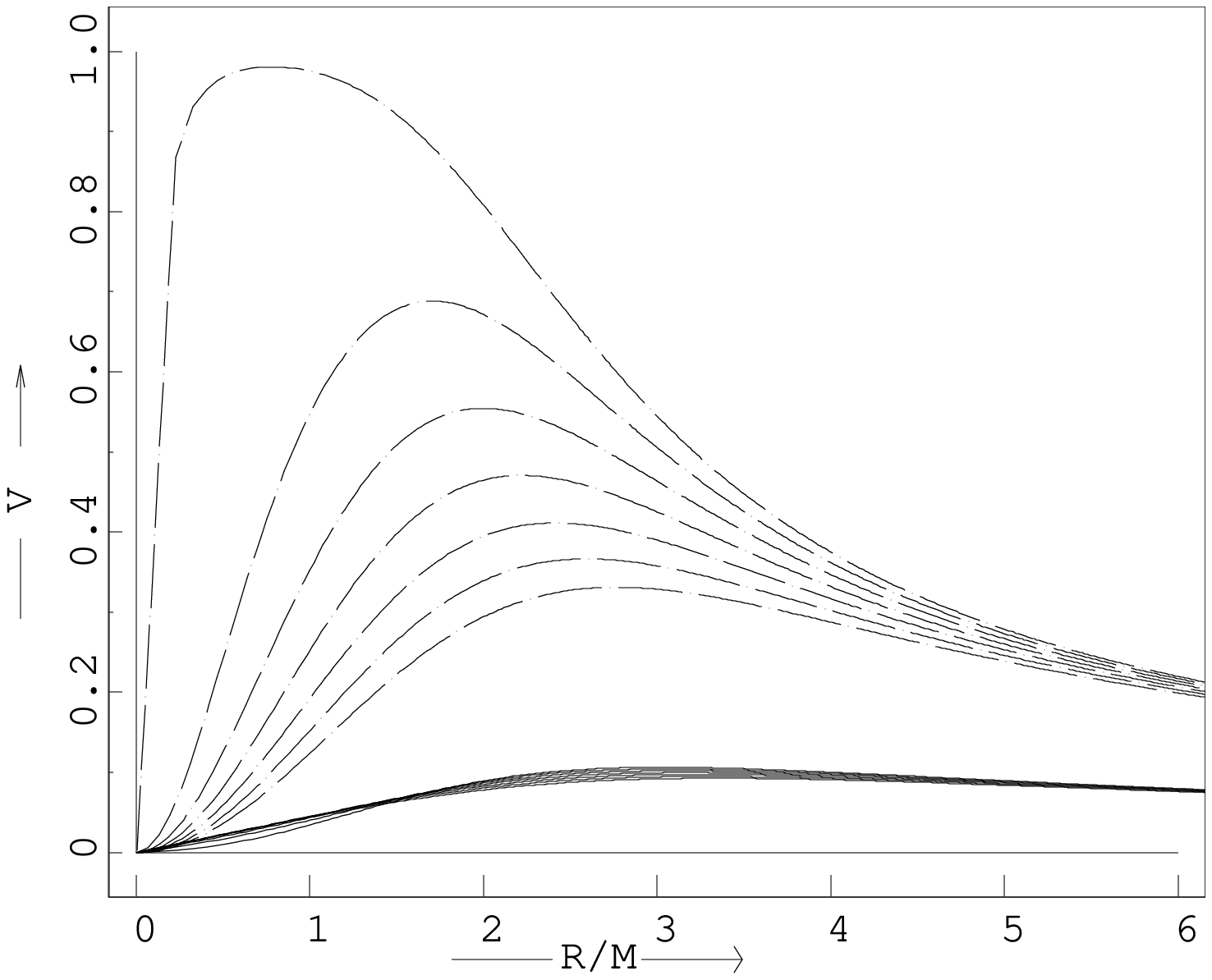,width=5.5cm}\\
Figure 1. Radial distribution of mass \>Figure 2. Velocity of FIO with respect \\
density as seen by FIO                \>to LNRF and the ratio of azimuthal\\
                                      \>pressure to mass density (dashed)\\
\psfig{file=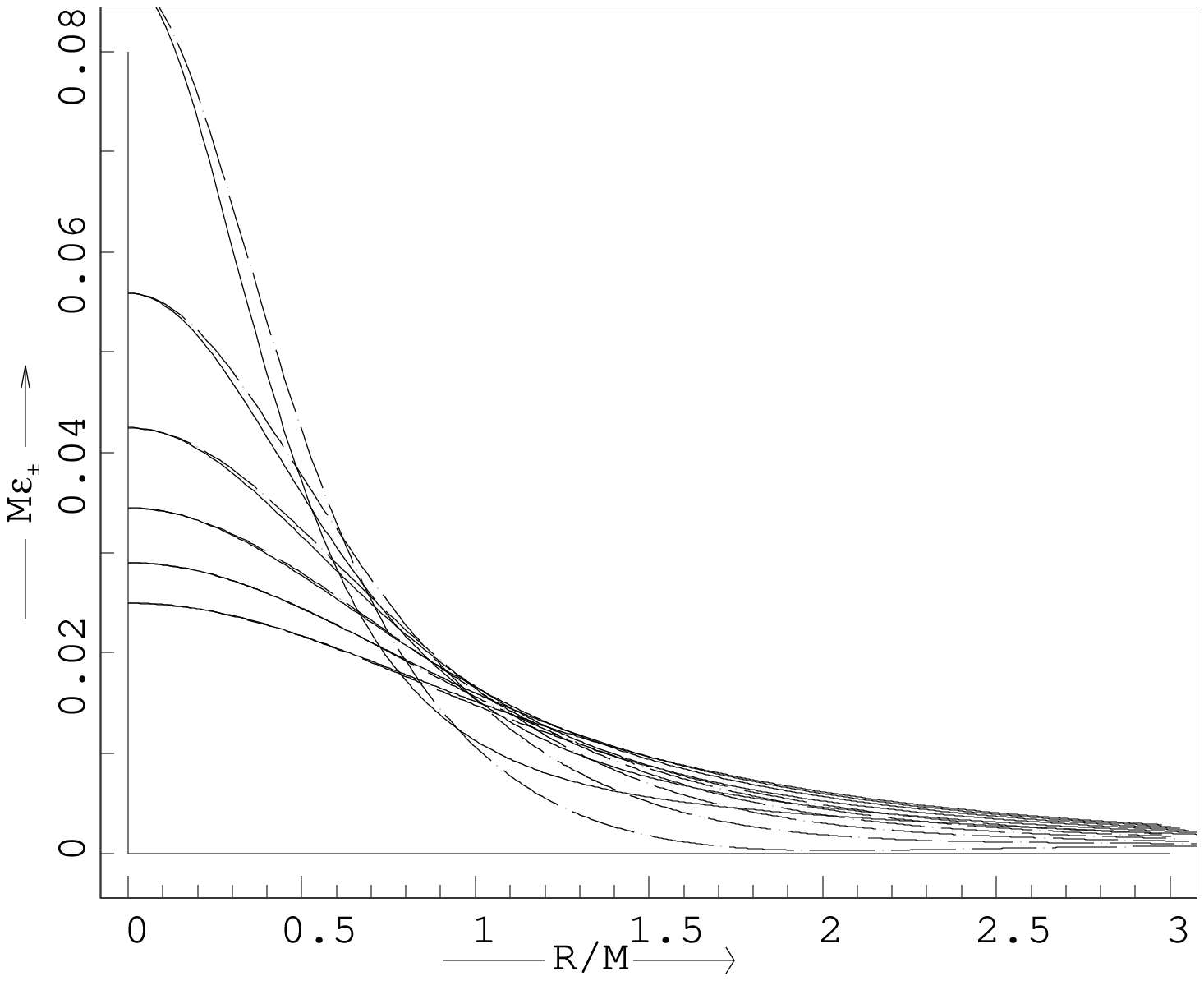,width=5.5cm}\>
\psfig{file=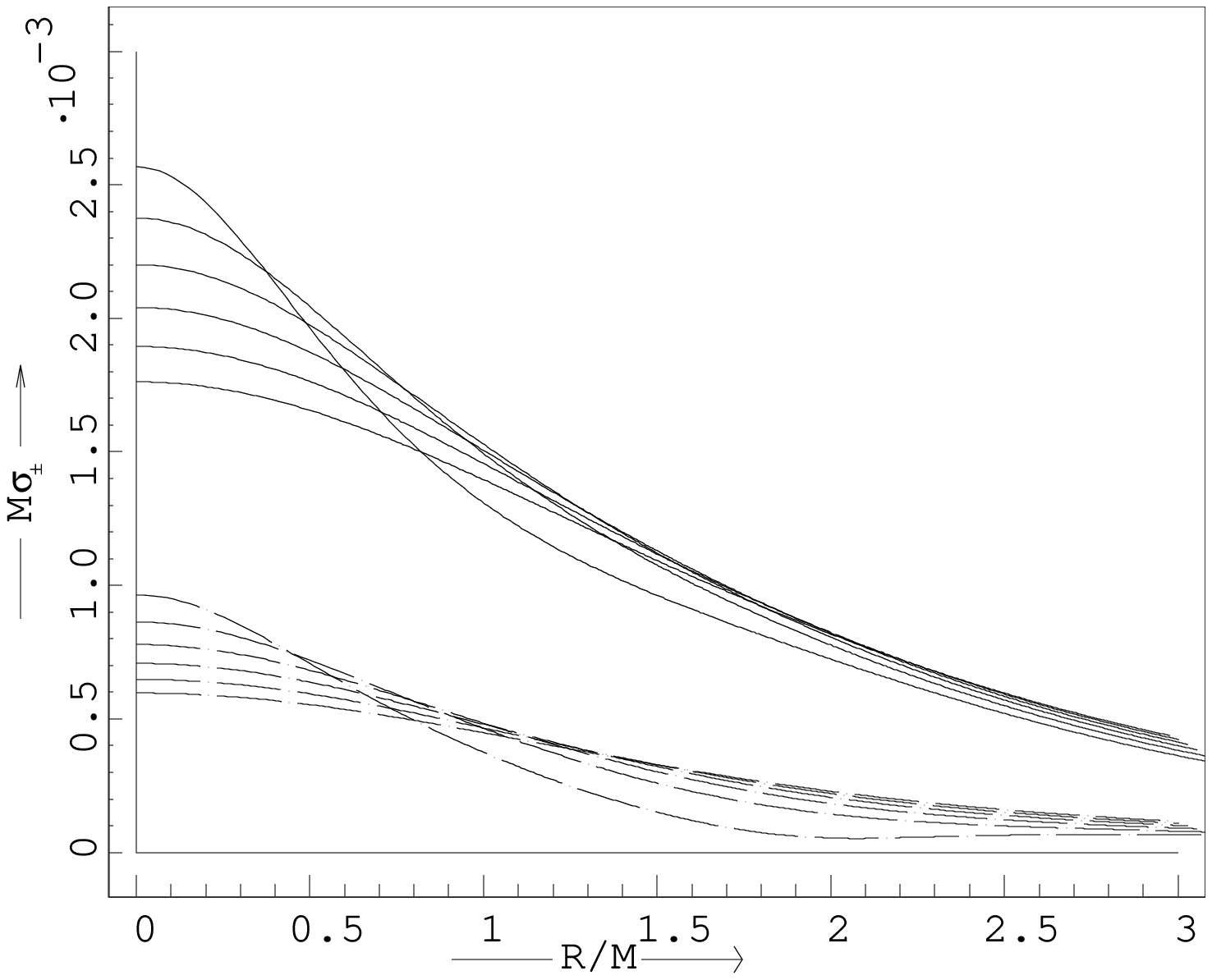,width=5.5cm}\\
Figure 3. Distribution of mass density\>
Figure 4. Distribution of charge density\\
in both streams $\epsilon_+$ and $\epsilon_-$ (dashed)\>
in both streams $\sigma_+$ and $\sigma_-$ (dashed)\\
\end{tabbing}

\end{document}